%% file: MAIN.tex
\documentclass[journal]{IEEEtran}
\usepackage{comment}
\usepackage{graphicx} % Essential for images
\usepackage{amsmath, amssymb, amsfonts} % Math packages, combine them
\usepackage{cite}     % For citations
\usepackage{rotating} % For rotating elements
\usepackage{pdflscape} % For landscape pages in PDF
\usepackage{booktabs}  % For professional tables
\usepackage{siunitx}   % For units and numbers in tables/text
\usepackage{mathtools} % Enhancements for amsmath
\usepackage{cuted}     % For cut environments (e.g., wide equations)
\usepackage{multirow}  % For multirow cells in tables
\usepackage{eqnarray}  % If you specifically need eqnarray, though amsmath is usually preferred
\usepackage{placeins}  % For \FloatBarrier
\usepackage{float}     % For better float control (e.g., [H] option)
\usepackage{lipsum}    % For dummy text
\usepackage{flushend}  % For balancing columns on last page (two-column layouts)
\usepackage{balance}   % Alternative/complement to flushend
\usepackage{scalerel}  % For scaling content
\usepackage{stfloats}  % For floats at bottom of two-column pages
\usepackage{bigints}   % For big integrals
\usepackage{textcomp}  % For extra text symbols
\usepackage{tabularx}  % For tables with flexible column widths
\usepackage{cleveref}  % For smart cross-referencing (load late!)
\usepackage{ragged2e}  % For justified text with hyphenation control
\justifying            % Apply justification
\usepackage{makecell}  % For line breaks in table cells
\usepackage{tcolorbox} % For colored boxes/frames
\usepackage{stackengine} % For stacking elements
\usepackage[export]{adjustbox} % For adjusting images/boxes
\usepackage{threeparttable} % For table notes
\usepackage{longtable} % For tables spanning multiple pages
\usepackage{breqn}     % For automatic line breaking in equations (can conflict sometimes, test carefully)
\usepackage{csquotes}  % Recommended for bibliography packages like biblatex
\usepackage{mwe}       % Minimal Working Example (often for examples, may not be needed in final doc)
\usepackage{url}       % For breaking URLs
\usepackage{algpseudocode} % Algorithm pseudocode (if you use this style)
\usepackage{algorithm}     % Basic algorithm float
\usepackage{setspace}      % Line spacing control
\usepackage{nomencl}       % For nomenclature
\usepackage{etoolbox}      % Required by nomencl for some features
\makenomenclature          % Command for nomenclature
\usepackage{acronym}       % For acronyms
\usepackage{array}         % Extensions for array/tabular environment
\usepackage{xcolor, color} % For colors (often part of graphicx, but explicit is fine)

\usepackage{subfig}
\usepackage{caption}
\begin{document}

\title{Diffraction and Scattering Modeling for Laser Power Beaming in Lunar Environment}

\author{Yanni~Jiwan-Mercier,~\IEEEmembership{Graduate Member,~IEEE,} Barış~Dönmez,~\IEEEmembership{Graduate Member,~IEEE,}       
        ~Güneş~Karabulut~Kurt,~\IEEEmembership{Senior~Member,~IEEE} and Sébastien~Loranger,~\IEEEmembership{Senior~Member,~IEEE}% <-this % stops a space
\thanks{Y. Jiwan-Mercier, B. Donmez, S. Loranger, and G. Karabulut Kurt are Poly-Grames Research Center, the Department of Electrical Engineering, Polytechnique Montréal, Québec, CA e-mail: yanni.jiwan-mercier@polymtl.ca}
}

% The paper headers
\markboth{Journal of \LaTeX\ Class Files,~Vol.~14, No.~8, July~2025}%
{Jiwan-Mercier \MakeLowercase{\textit{et al.}}: Height-Dependent Diffraction and Absorption Modeling for Laser Power Beaming in Lunar Dust}

% make the title area
\maketitle

% As a general rule, do not put math, special symbols or citations
% in the abstract or keywords.
\begin{abstract}
Reliable energy delivery is a critical requirement for long-term lunar missions, particularly in regions with limited solar access, such as polar craters and during extended lunar nights. Optical Power Beaming (OPB) using high-power lasers offers a promising alternative to conventional solar power, but the effects of suspended lunar dust on beam propagation remain poorly understood. This study introduces a detailed simulation model that incorporates both diffraction and height-dependent scattering by the electrostatically suspended lunar regolith. Unlike prior approaches, which assumed uniform dust layers or center-to-center transmission loss, our model uses generalized diffraction theory and refractive index gradients derived from particle density to assess beam deformation and attenuation. The results show that even in ground-to-ground scenarios, lunar dust significantly degrades energy transfer efficiency, dropping from 57\% to 3.7\% over 50 km in dust-free vs. dusty conditions with 175 nm particles. Increasing the particle size to 250 nm limits the viable transmission range to below 30 km at 6\% efficiency. The study further demonstrates that raising the laser source height can improve efficiency, achieving 91\% for a distance of 5 km and 25\% at 50 km when the source is positioned 12 m above ground. These findings underscore the importance of system elevation and dust modeling in lunar OPB design and reveal the mission-critical role of particle size distribution, especially in environments disturbed by human activity.
\end{abstract}

% Note that keywords are not normally used for peerreview papers.

%\IEEEpeerreviewmaketitle

%\section*{Nomenclature}
\label{Nomen}
%\begin{nomenclature}
\nomenclature[01]{($x$, $y$, $z$)}{Position of destination points (receiver panel)}
\nomenclature{$d_p$}{Diameter of lunar dust particle}
\nomenclature{$I(x,y,z)$}{Irradiance of the gaussian beam at point $(x,y)$ and at propagation distance $z$ on the propagation axis}

\nomenclature{($x_0$, $y_0$, 0)}{Position of spherical-source element}
\nomenclature{$(0,h_0,0)$}{Center position of Gaussian laser source}
\nomenclature{$(0,h_p,D)$}{Center position of receiver panel}
\nomenclature{$R$}{Target beam element propagation from source point to panel point}
\nomenclature{$N$}{Particle density [m$^{-3}$]}
\nomenclature{$n$}{Refractive index}

\nomenclature{$h$}{Height of the beam to the ground  [m]}

\nomenclature{$P_0$}{Power of the laser [W]}
\nomenclature{$E_0$}{Electric field at the aperture [V m$^{-2}$]}

\nomenclature{$\eta$}{Wave impedance [$\Omega$]}

\nomenclature{$\omega_0$}{Beam radius  at the waist [m]}
\nomenclature{$\omega(z)$}{Beam radius at distance $z$ from the waist [m]}
\nomenclature{$\lambda$}{Wavelength of the laser beam [m]}
\nomenclature{$\Phi$}{Phase change of wave of the beam}
\nomenclature{$P_r$}{Power received at the solar panel [W]}
\nomenclature{$L$}{Length of the panel (size along the $x$ axis) [m]}
\nomenclature{$W$}{Width of the panel (size along the $y$ axis) [m]}
\nomenclature{$C_{ext}$}{Extinction cross section [m$^2$]}
\nomenclature{$\theta$}{Angle of inclination of the laser and the panel}
\nomenclature{$D$}{Distance between the center position of the source and the center position of the panel [m]}
\printnomenclature
%\end{nomenclature}
%*****************%
%
%*****************%

%****************************************%
\section{Introduction}
\input{Sections/IntroLunarDust}
\label{Sec:intro}
%****************************************%
\subsection{Literature Review}
\input{Sections/LiteratureReview2}
\label{Sec:litreview}
%****************************************%
\subsection{Contributions}
\input{Sections/Contributions}
\label{Sec:contrib}
%****************************************%
\subsection{Organization}
\input{Sections/Organization}

\label{Sec:organization}
%****************************************%
%\section{Laser Beam Propagation Model}
%\input{Sections/LaserModel}
%\label{Sec:LaserMod}

\section{Diffraction Model}
\input{Sections/DiffractionModel}
\label{Sec:DiffracMod}
%****************************************%
\section{Lunar Dust Model}
\input{Sections/LunarDustModel}

\label{Sec:lunardust}
%****************************************%
\section{Numerical Results and Discussion}
\input{Sections/ResultsDiscussion}
\label{Sec:results}
%****************************************%
\section{Conclusion}
\input{Sections/Conclusion}
\label{Sec:conc}
%****************************************%

% ---- Bibliography ----
\bibliographystyle{IEEEtran}
\bibliography{References}

%****************************************%
%\section{Biography}
\input{Biography}

\label{Sec:bio}
%****************************************%

\end{document}

%% file: Sections/IntroLunarDust.tex
The strong global interest in lunar missions through NASA’s Artemis program, the European Space Agency, and private companies comes with a clear goal: establish a long-term presence on the Moon \cite{nasa_artemis,esa_moon,private_moon}. However, objectives like the exploration of the surface, mining operations, and the establishment of a Moon village all involve the deployment of a great number of devices, which all require reliable, sustainable power delivery \cite{nasapowerneeds}. 
The lunar far side (LFS) of the Moon receives a lack of sunlight for 14 days, and it limits the exploration missions; however, a power transfer from an external source can be a solution for the survival problem of the space vehicles operating on the LFS \cite{donmez2024continuous}.

One proposed solution is the use of a wireless power transfer (WPT) system in which a high-power laser targets the solar panel of these devices to supply energy when sunlight is inaccessible \cite{naqbi2024opticalpowerbeaminglunar}. Two main scenarios are typically considered; ground-to-ground and orbit-to-ground \cite{naqbi2024opticalpowerbeaminglunar}. In ground-to-ground power transfer, the laser source is positioned on a tower, while in the orbit-to-ground system, the laser is placed on a satellite in orbit around the Moon. The two solutions have their strengths and limitations, but both benefit from the unique conditions of the Moon’s environment. The scenario for which the current study is primarily aimed is ground-to-ground; however, the model that will be presented can also be used for orbit-to-ground systems for beams with a large vertical angle.

The lunar environment is often described as ideal for power beaming and telecommunication due to the absence of atmosphere \cite{naqbi2024opticalpowerbeaminglunar,Stubbs2006Dynamic}. Unlike the Earth, where atmospheric perturbations degrade optical transmission, the Moon’s surface and orbit are nearly equivalent to free space. The main factor that prevents this environment from being considered a perfect vacuum is the lunar regolith, which covers the surface of the Moon and can rise into suspension to create dust clouds \cite{horanyi2015}. A permanent dust cloud envelops the sun-exposed surface due to electrostatic charging from solar wind and UV radiation \cite{Stubbs2006Dynamic}. Particle suspension can also be caused by human activity or meteoritic impacts—from micro-meteoroids to larger events~\cite{szalay2016}.

Lunar dust negatively impacts both static optics (e.g., laser apertures and receiver panels) and beam propagation itself, primarily through scattering and absorption effects \cite{naqbi2024opticalpowerbeaminglunar}. Since lunar dust is generally concentrated within a few meters of the surface, and not in sufficient density to significantly affect short-range propagation, its effect is often neglected—especially in orbit-to-ground or short-range ground-to-ground systems \cite{naqbi2024opticalpowerbeaminglunar}. However, over very long distances (tens of kilometers), the cumulative scattering and attenuation effects become substantial and must be carefully modeled for WPT system design.

Although the interest for lunar wireless power transfer is flourishing, few studies give information on the impact of lunar dust particles on the performance of OPB systems, especially over longer transmission distances. Existing models usually neglect the effects of dust on power beaming, making it difficult to have references on the properties of dust at the surface of the Moon. This paper addresses this gap by simulating the lunar environment and its effects a laser beam for varying dust particles parameters. 

In the following subsection, we provide a comparative literature review to situate our work within the broader research landscape. That is followed by a summary of the paper’s main contributions and a brief outline of its structure.

%% file: Sections/LiteratureReview2.tex
Wireless energy harvesting systems in space have been gaining attention along with the technological advancements in the laser communication systems industry (e.g., solar cell efficiency). The lunar ground station establishes WPT to a spacecraft on the Moon in \cite{bozek1995laserbeam}. A laser diode with an 800 nm wavelength and 30\% efficiency is taken into consideration, together with the solar cells having an efficiency of 20\%. Multi-orbiter power beaming with various orbit configurations is investigated in \cite{donmez2025multiorbitercontinuouslunarbeaming}. Continuous power supply to a lunar rover located at the south pole can be established with a 40-satellite scheme during a revolution of the Moon around the Earth in case of no stationkeeping. In \cite{donmez2025hybrid}, the feasibility analysis of a multi-hop hybrid RF/FSO WPT system over the Moon is conducted. The harvested power results are compared for perfect alignment and Rayleigh distributed laser misalignment fading scenarios. Moreover, the Volta Space Technologies company successfully demonstrated autonomous tracking and laser charging of a lunar rover for more than 200 meters while staying above the energy threshold levels (i.e., for survival in the lunar environment)~\cite{VoltaSpace}.

%GENERAL to SPECIFIC --> LASER-BASED WPT in SPACE to Lunar DUST MODELS

%\textcolor{red}{Reduce the number of references/studies about laser-based WPT in SPACE.}

In 2002, photovoltaic efficiencies above 50\% were achieved under high-intensity illumination of 42 W/cm$^2$, highlighting the importance of thermal management \cite{VanRiesen2001Gaas}. A year later, NASA’s Marshall Space Flight Center powered a small aircraft with a 940 nm and 1.5 kW laser at a distance of 15 meters, with 17.7\% efficiency using Ga:In:P2 triple-junction cells \cite{Raible2008High}. In 2014, researchers demonstrated an OPB system transmitting 24 W at 793 nm over 100 meters with an 11.6\% electric-to-electric efficiency \cite{He2014HighPower}. The Japan Aerospace Exploration Agency achieved 74.7 W electrical power from a 350 W laser over 200 m, marking progress in OPB system precision and power delivery \cite{York2017High}. In 2019, the US Naval Laboratory demonstrated receiving 400 W from a 2 kW laser at 325 m \cite{Cavallaro2019Researchers}. All these experiments were conducted on Earth, where the atmosphere leads to major losses. Results in a space environment are expected to be even more efficient.

In 2015, a study reported 44\% efficiency at 4 kW output for high-brightness diode lasers, showing diode technology can still be optimized for high-power OPB \cite{Huang2015Recent}. Platonov \textit{et al.} demonstrated in 2020 a 1.33 kW ytterbium fiber laser at 1018 nm with beam quality of M$^2 < 1.1$, confirming fiber lasers' suitability for high-efficiency OPB \cite{Platonov2020HighEfficient}. Commercially available ytterbium fiber lasers operating between 1007–1070 nm match well with InGaAs laser power converters (LPC) and offer stable, high-efficiency power transmission \cite{IPG2025High}. This was demonstrated in 2020 with 50.8\% PCE for InGaAs-based LPCs from a 1064 nm laser \cite{Nikolay2020Optimization}. By matching the photon energy and the bandgap energy of the reviving material, Helmers and al. reported peak efficiencies of 68.9\% with 858 nm light \cite{Helmers202168}. A 2022 study by Fafard and Masson showed 74.7 \% laser power conversion efficiency with 808 nm in a low temperature environment (150 K) \cite{Fafard202274}. The 2023 SWELL experiment by the US Naval Research Laboratory successfully transmitted 1.5 W over 1.45 m in orbit for over 100 days, achieving 11\% efficiency in the longest and most powerful in-orbit OPB demonstration to date \cite{USNRL2023First}.

On the specific subject of power transfer through lunar dust, Naqbi investigated the feasibility of high-intensity OPB for powering lunar infrastructure, considering lunar dust interference, thermal dissipation, and size, weight, and power constraints \cite{Naqbi2024Impact}. The study developed comprehensive models incorporating Mie and T-matrix theory to quantify optical losses due to lofted lunar dust, and uses Gaussian beam theory to simulate beam propagation and energy harvesting. Simulation results indicate that dust can significantly attenuate near-surface transmission, but strategic system designs, such as elevated platforms and optimized apertures, can mitigate its effects. This demonstrates OPB’s potential as a reliable energy solution for long-duration lunar missions as well as a model to calculate lunar dust particles density as a function of altitude.

The Moon is covered by a cloud of electrostatically suspended lunar dust \cite{Rennilson1974Surveyor}. However, precise knowledge regarding this cloud remains a significant gap due to lack of data in the field. Indeed, most studies are models based on very scarce observations and analysis of few samples of lunar regolith brought back to Earth.
There are two main mechanisms to explain the presence of suspended dust particles above the surface: electrostatic levitation and micrometeoroid impact ejection.
Electrostatic levitation comes from sunlight charging lunar regolith when the surface is exposed, particularly at sunrise and sunset when strong horizontal electric fields can develop due to differential surface charging, which causes submicron particles to levitate into transient suspension \cite{Colwell2007Lunar}. It was suggested by Stubbs \textit{et al.} that charged particles are launched upward and follow ballistic trajectories before returning to the surface \cite{Stubbs2006Dynamic}. Experimental studies have confirmed that vertical electric fields on the order of 10 V/m are sufficient to lift submicron dust grains under simulated lunar conditions \cite{Colwell2007Lunar}.
Micrometeoroid impacts also contribute to dust lofting as it can propel upward regolith grain lying on the surface \cite{Poppe2010Simulations}. LADEE's Lunar Dust Experiment (LDEX) confirmed the presence of impact-ejected dust clouds around the Moon \cite{Szalay2015Search}.

In terms of size, lunar dust particles are usually small and irregularly shaped \cite{Agui2007Lunar} and their size ranges from nanometers to millimeters \cite{Vidwans2022Size}. Grains lifted because of electrostatic levitation are generally less than 1 $\mu$m in diameter \cite{Hartzell2011Role}. Once they are lofted, it was shown that particles as small as 0.07 $\mu$m can remain stably suspended in illuminated regions, while larger particles will return to the surface quickly because of the photoelectron sheath at the surface \cite{Poppe2010Simulations}.
In terms of morphology, dust particles are highly angular and jagged, unlike rounded grains found on Earth’s surface \cite{Gaier2005Effects}. They come from rock fragments, glass beads and agglutinates formed by micrometeoroid impacts. And they can also be coated with smaller iron particles in the nanometer range due to space weathering, which contributes to high abrasiveness and unique electrical properties \cite{Taylor2001Lunar}.
Observations by the LADEE mission indicated that the density of 0.3 $\mu$m particles decreases exponentially with altitude \cite{Szalay2016Lunar}. 20 km to 100 km above the ground, typical densities were in the range of 10$^2$ to 10$^3$ particles per cubic meter \cite{Horanyi2014Lunar}. Closer to the surface, densities could reach up to 10$^5$ particles/m$^3$ during impact events or near human activity \cite{Jin2024Properties}.

We can draw from the literature that the existence of a dust cloud above the surface of the Moon is known and the mechanisms explaining the density, compositions and other properties can be theorized and explained. However, the accurate characteristics of lunar dust cannot be properly understood because there is not sufficient direct data from the Moon environment. We rely mainly on limited samples brought back to Earth to run experiments. The realistic behavior of lunar dust in various conditions like illumination, altitude or human activity cannot be fully characterized. 

While LADEE provided high-altitude data \cite{Szalay2015Search}, the challenge now is to obtain more measurements from the surface dust environment. Models of the lunar dust particle density and the concentration of different particle sizes as functions of the altitude are missing.

%% file: Sections/Contributions.tex
This paper presents rigorous simulations to quantify the impact of lunar dust on ground-to-ground OPB, with different particle size diameters to account for the absence of accurate data on the subject. While previous models have either assumed simplified uniform attenuation or ignored dust effects, this study incorporates generalized diffraction theory and height-dependent refractive index variations caused by the distribution of lunar dust particles at the surface of the Moon. The key contributions of this work are:

\begin{itemize}
    \item Development of a laser diffraction model that simulates laser beam propagation through lunar dust with a height-dependent complex refractive index.
    \item Analytical modeling of the lunar dust density distribution and its effect on the real and imaginary parts of the refractive index as a function of height.
    \item Quantitative analysis of beam deviation, scattering attenuation, and irradiance deformation over various long distances.
    \item Power transfer performance evaluation for different particle sizes, distances between source and receiver, and source heights, showing how efficiency drops with increased dust size, distance and proximity to the ground. 
\end{itemize}

%% file: Sections/Organization.tex
The remainder of this paper is organized as follows:  Section~\ref{Sec:DiffracMod} introduces the diffraction model used for beam propagation and to calculate received power. Section~\ref{Sec:lunardust} describes the lunar dust model, including the particle density distribution as a function of height and the impact of refractive index on the phase of the laser beam. Section~\ref{Sec:results} presents simulation results and discusses the influence of distance, source and receiver heights, and particle size on power transfer. Section~\ref{Sec:conc} concludes the work and proposes directions for future studies.

%The remainder of this paper is organized as follows. In Section II, eight different satellite configurations, which are single, double, triple, and quadruple orbits with 30 and 40 LLO satellites, are evaluated based on the continuity during a Moon revolution around the Earth. In Section III, the laser-based WPT model is presented for the solar panel with and without the tracking ability. The impact of the cosine effect is illustrated, and the satellite selection method is introduced. In Section IV, the harvested power and overall system efficiency performance metrics are computed first, and then the results are compared between our benchmark model, which is 40-satellite quadruple orbit, and 30-satellite and a single satellite schemes for the two types of solar panels. The average values of the metrics over a Moon revolution allow us to evaluate the trade-off options. Finally, our work is concluded in Section V.}

%% file: Sections/DiffractionModel.tex
\begin{figure*}[ht]
\centering
\includegraphics[width=0.75\textwidth]{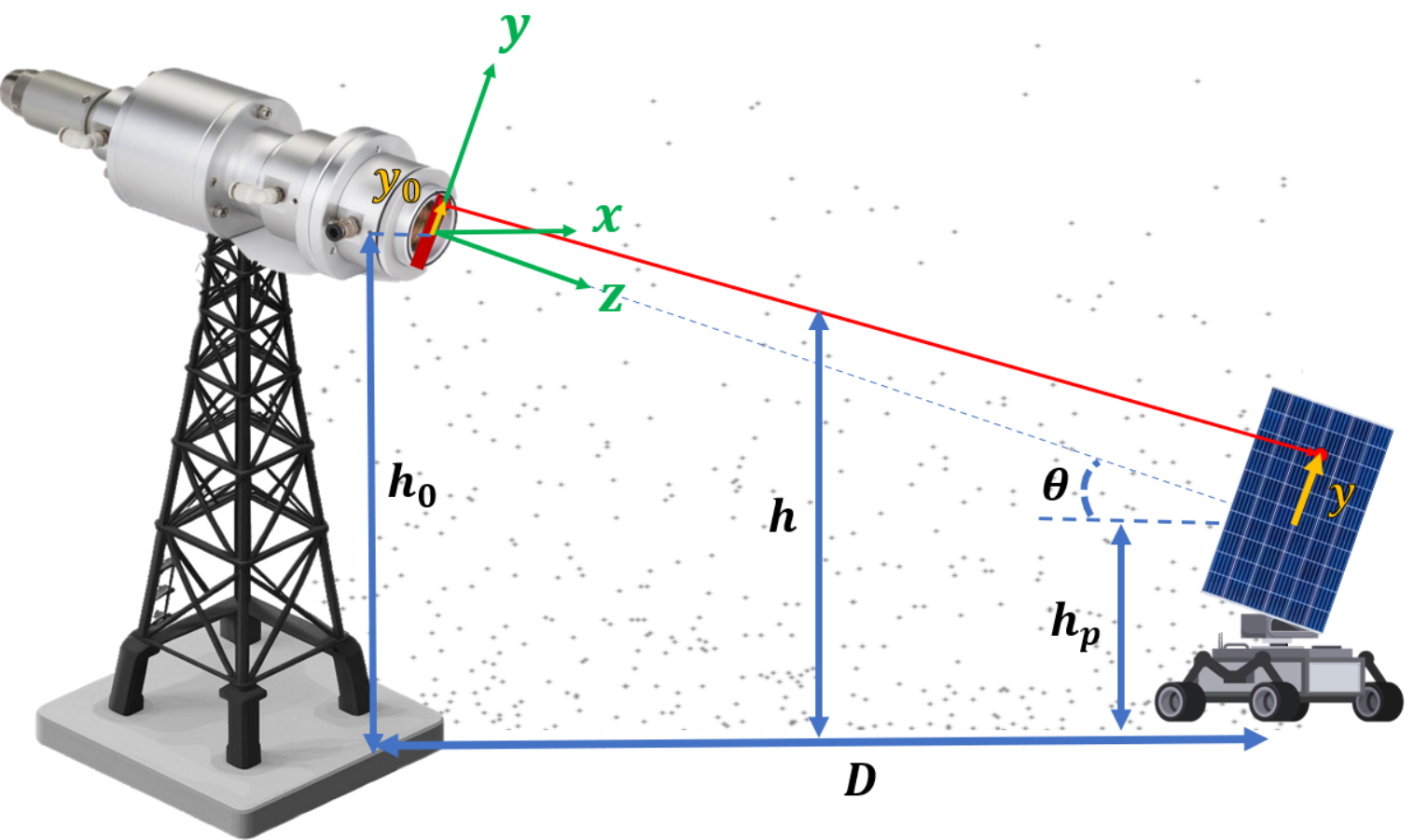}
\caption{\justifying{Illustration of the path of the laser from the source to the panel: the laser source is placed on a tower at height $h_0$, the beam propagates through lunar dust at height $h$ and reaches at distance $D$ the solar panel of the rover, whose center is at height $h_p$ and is aligned with the center of source which is the origin of the reference frame.}}
\label{fig:Sketch_laser_panel}
\vspace{-0.2cm}
\end{figure*}
Beam propagation is simulated using generalized diffraction, which is to consider the emitted field at the output of the laser as a distribution of infinitesimal individual spherical sources. The emitted laser output considered in this paper is a truncated Gaussian field perfectly collimated. Perfect collimation is implemented by considering the beam as a plane wave (all-in-phase) at the minimum diameter of divergence, which is called the waist. Since the laser source is aimed toward a receiving photovoltaic panel, we will consider a tilted reference frame where the z-axis is along the optical axis between the laser and panel centers, as depicted in Fig. \ref{fig:Sketch_laser_panel}. Once the beam is modeled at any arbitrary point in space, we will be able to integrate the power received by the panel. 
\begin{figure}
    \centering
    \includegraphics[width=1\linewidth]{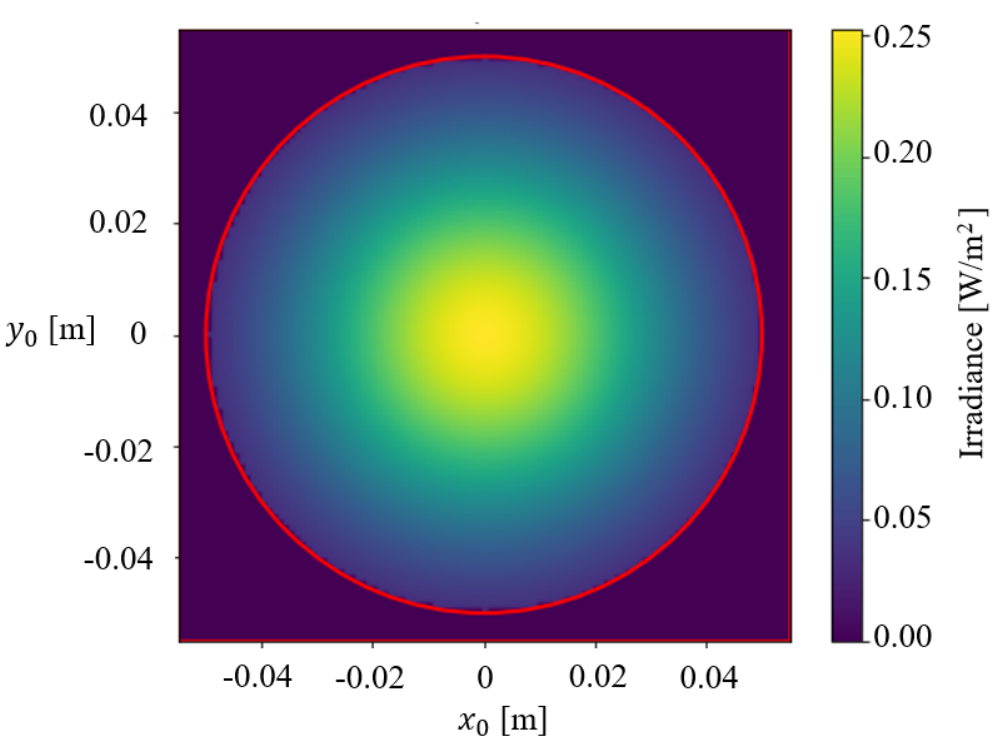}
    \caption{Gaussian distribution of the irradiance of the laser in the aperture (red circle) at the waist of the source. }
    \label{fig:irradiance_waist}
\end{figure}

To do so, let us first consider the emitted electric field of the laser beam in the aperture defined along the coordinates $(x_0,y_0,0)$ \cite{saleh2019fundamentals}:
\begin{equation} \label{eq:E_1source}
    E(x_0,y_0,0)= E_0(x_0,y_0) = \zeta \sqrt{\frac{4P_0\eta}{\pi \omega_0^2}}e^{\frac{-(x_0^2 + y_0^2)}{\omega_0^2}} g(x_0, y_0),
\end{equation}
\noindent where $P_0$ is the total output power of the laser, $\eta$ is the wave impedance (in free space, $\eta=377\Omega$) and $\omega_0$ is the Gaussian radius of the laser beam, defined at $1/e^2$ intensity from the maximum and is considered the beam waist. $g(x_0, y_0)$ is a window function equal to 1 for $x_0^2+y_0^2 < r_a$ (inside the aperture) and 0 otherwise, where $r_a$ is the aperture diameter truncating the Gaussian tail. Since the Gaussian is truncated, parameter $\zeta = \sqrt{1/(1-e^{-2})}$ is a constant added to keep the total output power through the aperture at $P_0$ in the case of an aperture radius equal to the waist. Fig. \ref{fig:irradiance_waist} shows the irradiance distribution of the source in the aperture. 

To model generalized diffraction, we consider the electric field from an infinitesimal point source element of the field of Eq. \eqref{eq:E_1source} at a point in space of coordinates $(x,y,z)$ \cite{saleh2019fundamentals}:
\begin{equation} \label{eq:dE}
 dE(x,y,z) = \frac{1}{\lambda R} E_0(x_0,y_0)~e^{-j\Phi} ~dA,
\end{equation}
where $dA$ is a small surface element of the source, $\lambda$ is the wavelength and $R$ is the distance between a source points $(x_0,y_0,0)$ and an arbitrary point in space $(x,y,z)$, calculated as:
\begin{equation}
     R = \sqrt{(x-x_o)^2 + (y-y_0)^2 + (z-z_0)^2}.
 \end{equation}

The component $\Phi$ is the the cumulative complex phase of the spherical wave between these two points. Since the complex refractive index $n$ has varying real and imaginary parts, the cumulative phase $\Phi$ is integrated over the optical path from the source to the arbitrary point in space \cite{saleh2019fundamentals}:
\begin{equation} \label{eq:phaseintegral}
     \Phi(x_0,y_0,x,y,z) = \int_{0}^{R} \frac{2\pi n(h(R'))}{\lambda}  \, dR',
 \end{equation}
where $n(h)$ is the complex refractive index and $h(R')$ is the height of the beam to the ground level, defined as:
\begin{equation}
    h(R') = \frac{(y-y_0)\cos\theta + h_p - h_0}{R}R' + y_0 \cos\theta + h_0,
\end{equation}
where $h_p$ and $h_0$ are the height of the center point of the panel and source, respectively. $\theta = \arctan\left(\frac{h_p - h_0}{D}\right)$ is the horizontal inclination between the source and the panel, as depicted in Fig. \ref{fig:Sketch_laser_panel}. The real part of $\Phi$ and $n$ represents the phase and the imaginary part represents the losses. The electric field at coordinates $(x,y,z)$ is obtained by integrating all the contributing sources of the aperture \cite{saleh2019fundamentals}:
\begin{equation}
    E(x,y,z) = \int_{}^{} \int_{}^{} \frac{1}{\lambda  R} E_0(x_0,y_0)~e^{-j\Phi} dy_0 dx_0,
\end{equation}
while the irradiance of the field at the point $(x,y,z)$ is then given by \cite{saleh2019fundamentals}:
\begin{equation}\label{eq:Irradiance}
    I(x,y,z) = \frac{E(x,y,z)E^*(x,y,z)}{2\eta},
\end{equation}
where $E^*(x,y,z)$ is the complex conjugate of $E(x,y,z)$.
In a uniform environment (i.e. non-varying index n), the phase becomes constant, and the irradiance calculated from Eq. \eqref{eq:E_1source} to \eqref{eq:Irradiance} is expressed as the classical Gaussian propagation \cite{saleh2019fundamentals}: %Fiber lasers propagate in a near-Gaussian mode, allowing them to be accurately modeled using Gaussian beam theory for free-space OPB applications%
\begin{equation}
I(x,y,z)=\frac{2\,P_0}{\pi w{{(z)}^{2}}}\exp \left( \frac{-2{({x}^{2} +y^2)}}{w{{(z)}^{2}}} \right) ,
\label{eq:irradiance_freespace} 
\end{equation}
where $\omega(z)$ is the beam radius (defined at $1/e^2$ intensity from maximum) at distance $z$ from the source defined as \cite{saleh2019fundamentals}: 
\begin{equation}
    \omega(z) = w_0 \sqrt{1-\left(\frac{z}{z_R}\right)^2},
    \label{eq:BeamRad}
\end{equation}
where $z_R = \frac{\pi \omega_o^2}{\lambda}$ is the Rayleigh range of the laser.

\indent In the general case of a non-uniform environment resulting from the presence of lunar dust which varies significantly with height, Eq. \eqref{eq:irradiance_freespace} and \eqref{eq:BeamRad} are no longer valid and the irradiance may suffer deformation, deviation and losses. 
At the rover, the beam is collected by a photovoltaic panel of length $L$ and width $W$ for which its center is at position $(0,0,D)$ in our tilted coordinate system.  The collected power is calculated by integrating Eq. \eqref{eq:Irradiance} over the surface of the panel:
\begin{equation}
    P_r  = \int_{-\frac{L}{2}}^{\frac{L}{2}} \int_{-\frac{W}{2}}^{\frac{W}{2}} I(x,y,z) dy dx.
\label{eq:power_integral}
\end{equation}
Due to asymmetry induced by the non-uniform index variation in the y-axis, Eq. \eqref{eq:power_integral} will be calculated numerically by using Eq. \eqref{eq:E_1source} to \eqref{eq:Irradiance}.

%% file: Sections/LunarDustModel.tex
The determining factor for measuring the impact of lunar dust is height: as the altitude increases, the particle density decreases. The model used to describe the density of lunar dust particles per cubic meter follows a logarithmic function~\cite{naqbi2024opticalpowerbeaminglunar}:
\begin{equation}\label{eq:ParticleDensity}
    N(h)=-4.166 \cdot 10^8 \ln{\frac{h}{8.68}}
\end{equation}
where $h \le 8.68$ is the height in meters.

\begin{figure}
    \centering
    \includegraphics[width=1\linewidth]{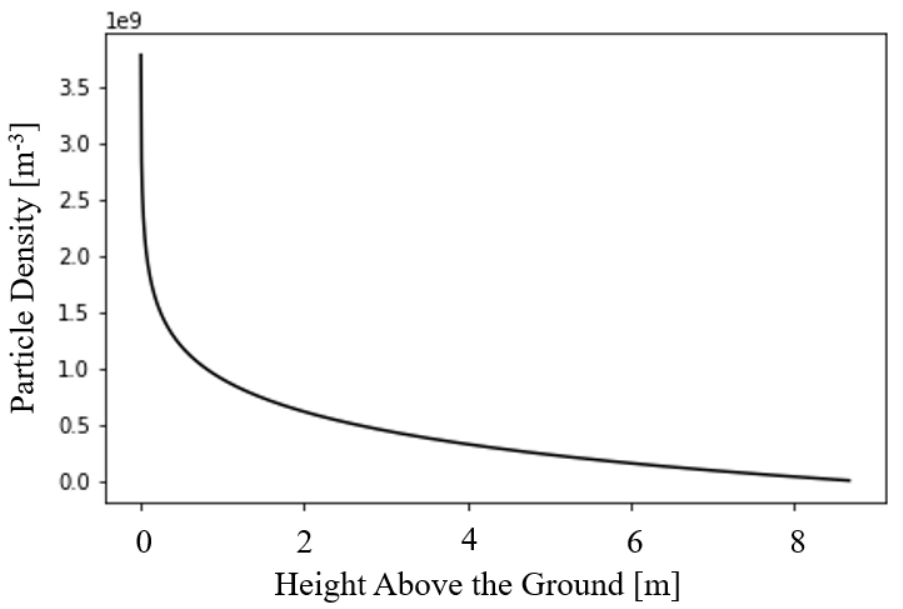}
    \caption{\justifying{Naqbi's Lunar dust particles density model following Eq. \eqref{eq:ParticleDensity}~\cite{naqbi2024opticalpowerbeaminglunar}}}.
    \label{fig:ParticleDensity_vs_Height}
\end{figure}

Fig. \ref{fig:ParticleDensity_vs_Height} shows Eq. \eqref{eq:ParticleDensity}. Over $8.68$ m, the model considers that there are negligible particles. The presence of these particles will create a gradient in the real-part index (i.e., the phase of the wave) viewed by the light and cause scattering losses which will be represented by an effective imaginary index.
To determine the real part of the index $n(h)$ as a function of height, we can neglect multiple reflections (localization effects from multiple scatterers) and consider propagation in a uniform medium, since the concentration is relatively small, and the particles are much smaller than wavelength. The index of this uniform medium can be approximated as the average volume ratio of dust particles: 
\begin{equation} \label{eq:refindex_re}
    \Re  (n(h)) = (1.733-1) N(h) \left(\frac{d_p}{2}\right)^3 \frac{4 \pi}{3} + 1,
\end{equation}
where $d_p$ is the particle diameter. Over the height of 8.68 m, the number of particles is considered negligible, therefore, the refractive index is 1.

For the imaginary part of the refractive index $n(h)$, it is calculated by multiplying the scattering cross-section $C_{ext}$ with the particle density function \cite{naqbi2024opticalpowerbeaminglunar}:
\begin{equation} \label{eq:refindex_im}
    \Im(n(h)) =\frac{C_{ext} N(h) \lambda}{2 \pi}.
\end{equation}
Over 8.68 m, the imaginary refractive index is considered negligible because of the absence of particles beyond this altitude in the density distribution model.

Fig. \ref{fig:Sketch_laser_panel} illustrates the propagation of the beam through the dust cloud from one point on the source to one point on the panel. The height-dependent real-part effective index causes refraction of the beam (i.e., deviation) towards the ground. For particles sizes of 175 nm, the deviation becomes noticeable at a range over 40 km when neglecting scattering losses. The heigh-decreasing scattering loss causes extinction of the lower-part of the beam, thus displacing the maximum point of the beam away from the ground. As can be observed in Fig. \ref{fig:IrradiancePatterns}, the effect of upward deviation indicates a dominance of scattering losses over refraction. Nevertheless, the refraction effect contributes to increasing losses, since it deviates the beam slightly towards the high-loss region.

Substituting Eq. \eqref{eq:refindex_re}-\eqref{eq:refindex_im} into Eq. \eqref{eq:phaseintegral}, we can obtain an analytical form of the cumulative phase: 
\begin{equation}
    \Re(\Phi) =\frac{2\pi P~R}{\lambda (y-y_0)}~ [y\ln{(B~y)} - y_0\ln{(B~y_0)} ] + R,
\end{equation}
where $ P = -4.166 \cdot 10^8 (1.75-1)(\frac{D_p}{2})^3 \frac{4 \pi}{3} $ and $B = 8.68$, and 
\begin{equation}
    \Im(\Phi) = C_{ext} ~ \frac{Q~R}{y-y_0}~ [y\ln{(B~y)} - y_0\ln{(B~y_0)} ] ,
\end{equation}
where $ Q= -4.166 \cdot 10^8 $ and $B = 8.68$\footnote{{When simulating,  a constant refractive index is considered when $y = y_0$, to avoid dividing by zero. 
}}.

%% file: Sections/ResultsDiscussion.tex
\begin{figure}
\centering
\subfloat[]{
	\label{subfig:centertocenter0175}
	\includegraphics[clip, scale=0.55]{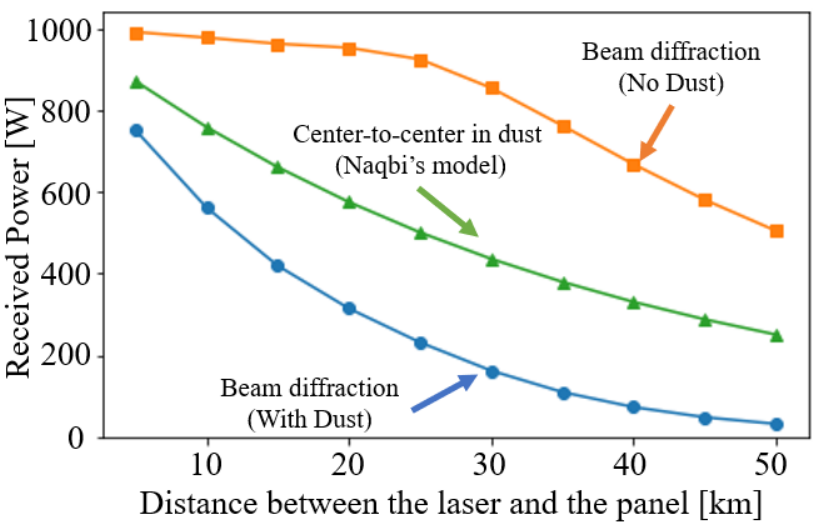}
	 }
     
\medskip
\subfloat[]{
	\label{subfig:centertocenter0176}
	\includegraphics[clip, scale=0.55]{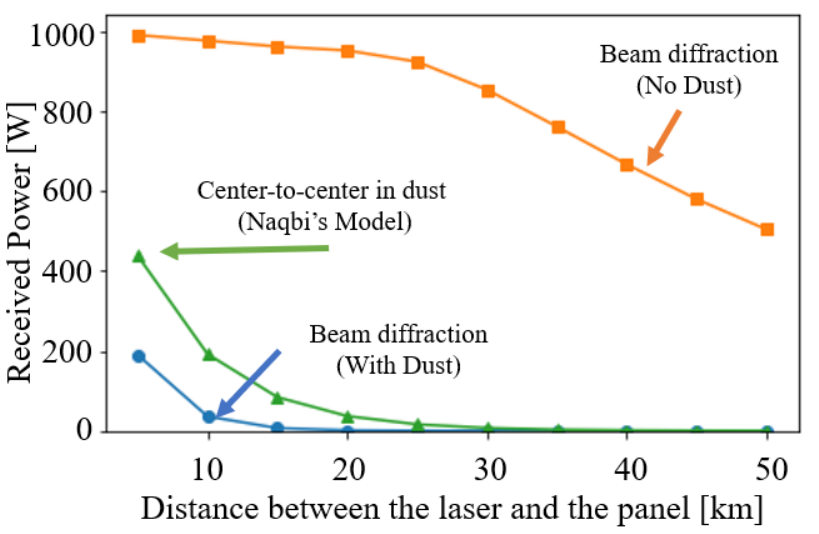}
	 }
\caption{ \justifying{Power received at the panel from a laser source sending 1 kW, both positioned 2 meters above the ground, through a dust cloud of particles of size (a) 0.175 $\mu$m and (b) 0.25 $\mu$m for 3 cases: \\
    GREEN: Single source in the center of the laser aperture propagating to a single point at the center of the panel. This is following Naqbi's dust model \cite{naqbi2024opticalpowerbeaminglunar} 
    \\ORANGE: Beam propagation following our diffraction model and propagating through free-space (no dust particles). 
    \\BLUE: Beam propagation following our diffraction model and propagating through lunar dust particles. }}
\label{fig:CenterToCenter_NoDust_Dust}
\vspace{-0.15cm}
\end{figure}
\begin{figure}
\centering
\subfloat[]{
	\label{subfig:2d0175}
	\includegraphics[clip, scale=0.6]{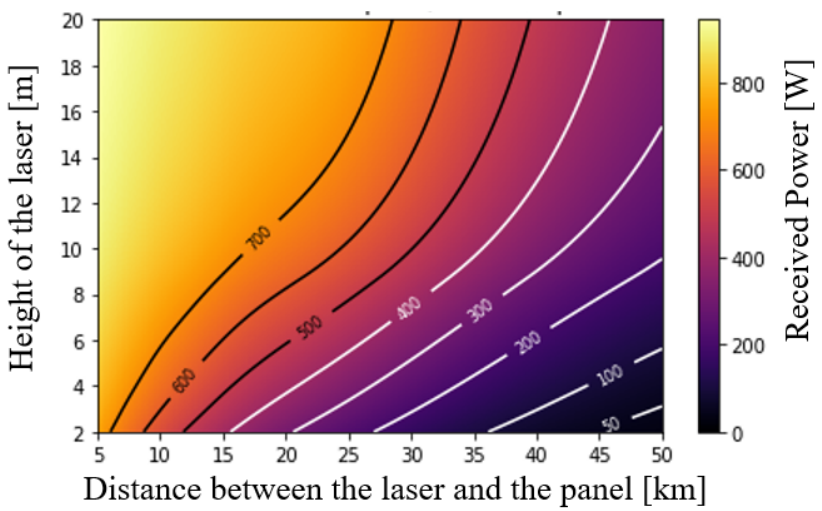}
	 }
     
\medskip
\subfloat[]{
	\label{subfig:2d025}
	\includegraphics[clip, scale=0.6]{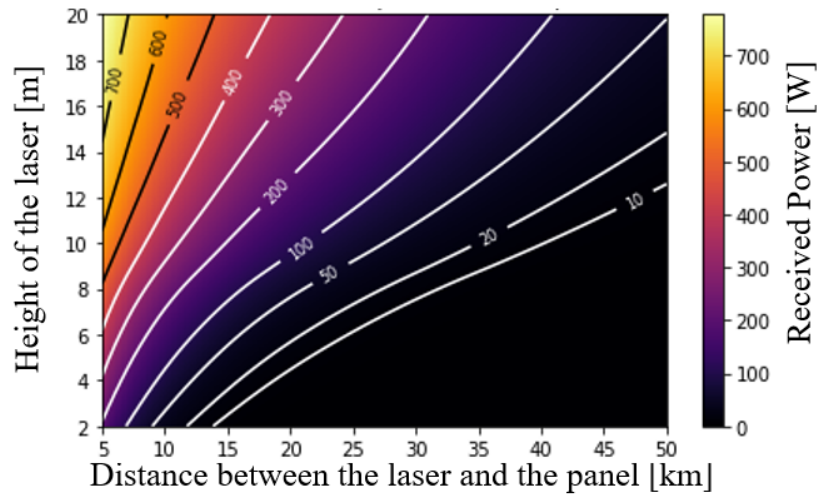}
	 }
\caption{\justifying{Received power at the panel when the beam propagates through dust for various heights of the laser source emitting 1 kW and various distances between the source and the solar array with particle diameter of (a) 0.175 $\mu$m and (b) 0.25 $\mu$m}}
\label{fig:PowerTransfer_2D_plot}
\vspace{-0.15cm}
\end{figure}
\begin{figure}
    \centering
    \includegraphics[width=1\linewidth]{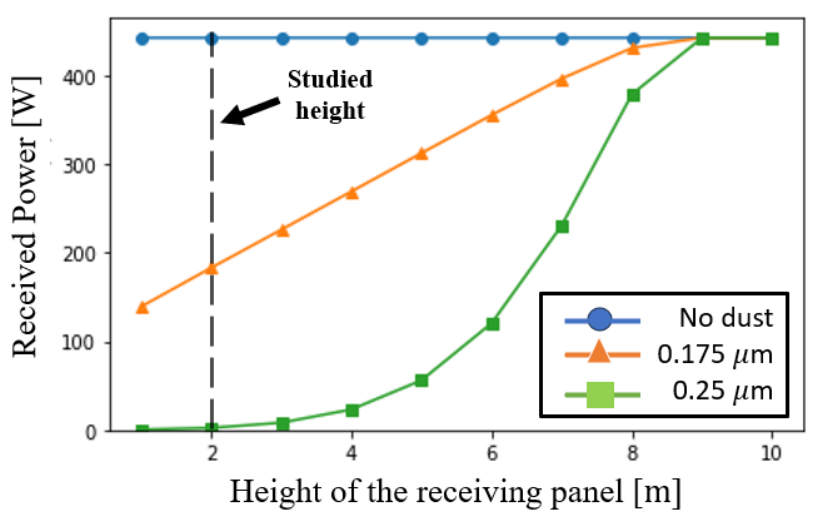}
    \caption{\justifying{Received power at the panel for fixed source height at 10 m and transmitting power of 1 kW, and increasing height of the receiving panel located at a distance of 50 km, in 3 scenarios: no dust and particles diameters of 175 nm and 250 nm. The panel height used in simulations is identified by the dashed line.}}
    \label{fig:PowervsHeightPanel}
\end{figure}

\begin{figure}
    \centering
    \includegraphics[width=1\linewidth]{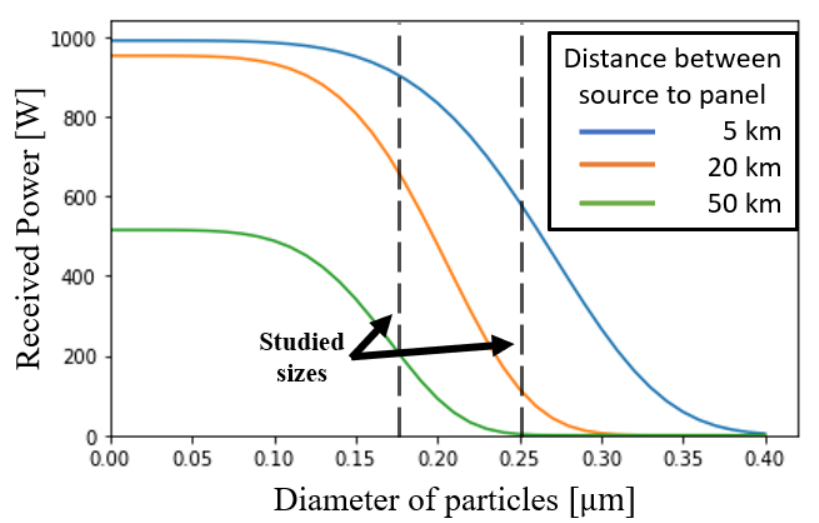}
    \caption{\justifying{Received power at the panel for increasing particle size diameter and different distances between source transmitting 1 kW and panel. A diameter of 0 means there is no dust and the beam propagates in free space. The two main particle size diameter studied in simulations are identified by the dashed line.}}
    \label{fig:PowervsParticle}
\end{figure}

Simulations were conducted with varying distances and source height with the parameters presented in Table \ref{table:simulations parameter}. The considered line-of-sight distances range from 0 km to 50 km and are evaluated to explore the performance limits of the architecture. Particles sizes of 175 nm were considered as the average suspended particles on the sun-exposed region of the Moon \cite{naqbi2024opticalpowerbeaminglunar}. Since the accuracy of this assumption is rather poor due to lack of data, we also consider particle slightly larger particles of 250 nm suspended with the same varying concentration to give an idea of the variability of transmission with particle size.
\\
\begin{table}[!t]
\vspace{0.05in}
  \sisetup{group-minimum-digits = 4}
  \centering
  \caption{Simulation Parameters}
  \label{table:simulations parameter}
  \begin{tabular}{lllS[table-format=5]ll} 
    \toprule
    \toprule
Transmit power ($P_0$)                                                     & 1 kW %\cite{donmezatp} 
\\ 
Laser wavelength ($\lambda$)                                                 & 1064 nm \cite{donmezatp}                               \\ 
% Beam divergence angle (${{\eta }_{T}}$)                                                      & Adaptive  \cite{donmezatp}                              \\ \hline
Beam waist (${{\omega}_{0}}$)                                                                 & 5 cm                           \\ 
%\begin{tabular}[c]{@{}l@{}}EMLP-2 stable satellite pointing \\ accuracy with ATP (${{\sigma }_{\gamma }}$) \end{tabular} & 5 nrad \cite{lasertypes}                                            \\ \hline
\begin{tabular}[c]{@{}l@{}}Laser telescope aperture diameter \end{tabular}     & 10 cm \cite{lasertypes}
\\ 
\begin{tabular}[c]{@{}l@{}} Diameter of particles ($d_p$)\end{tabular} & $0.175 ~\mu$m and $0.25 ~\mu$m        \\
\begin{tabular}[c]{@{}l@{}} Length of panel ($L$)\end{tabular} & 50 cm        \\
\begin{tabular}[c]{@{}l@{}}  Width of panel ($W$)\end{tabular} & 50 cm        \\
\begin{tabular}[c]{@{}l@{}}Height of the solar array ($h_p$) \end{tabular} & 2 m        \\

   \bottomrule
    \bottomrule
\end{tabular}
\end{table}

\begin{figure*}[!t]
\centering
\subfloat[]{
	\label{subfig:5km_0175um}
	\includegraphics[clip, scale=0.5]{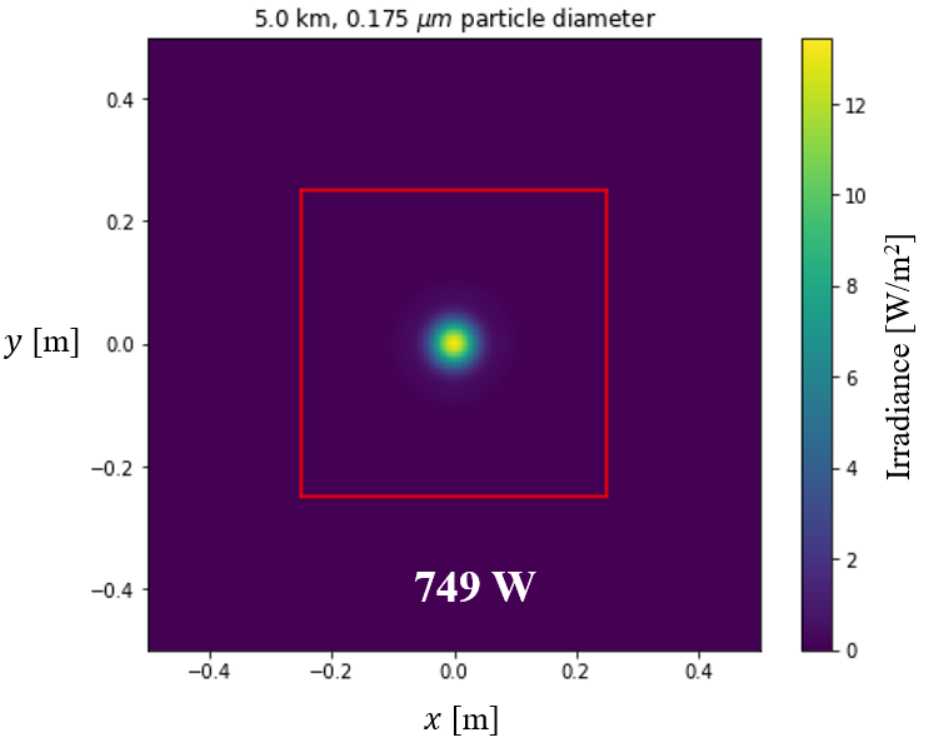}
	 }
\hfill
\subfloat[]{
	\label{subfig:5km_025um}
	\includegraphics[clip, scale=0.5]{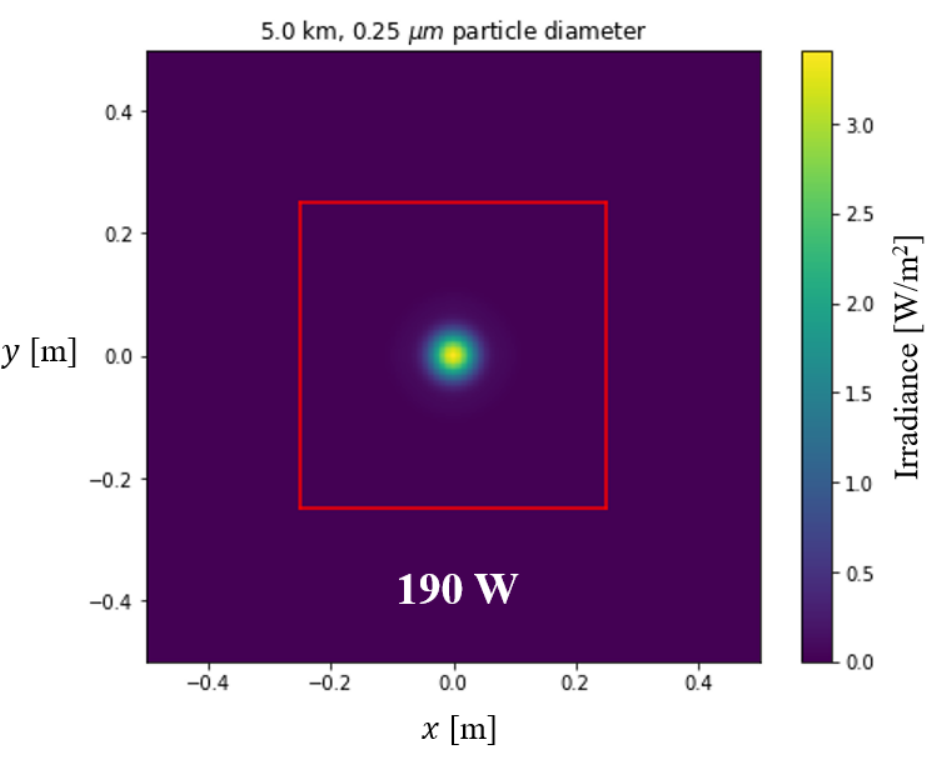}
	 }
     
\medskip
\subfloat[]{
	\label{subfig:20km_0175um}
	\includegraphics[clip, scale=0.5]{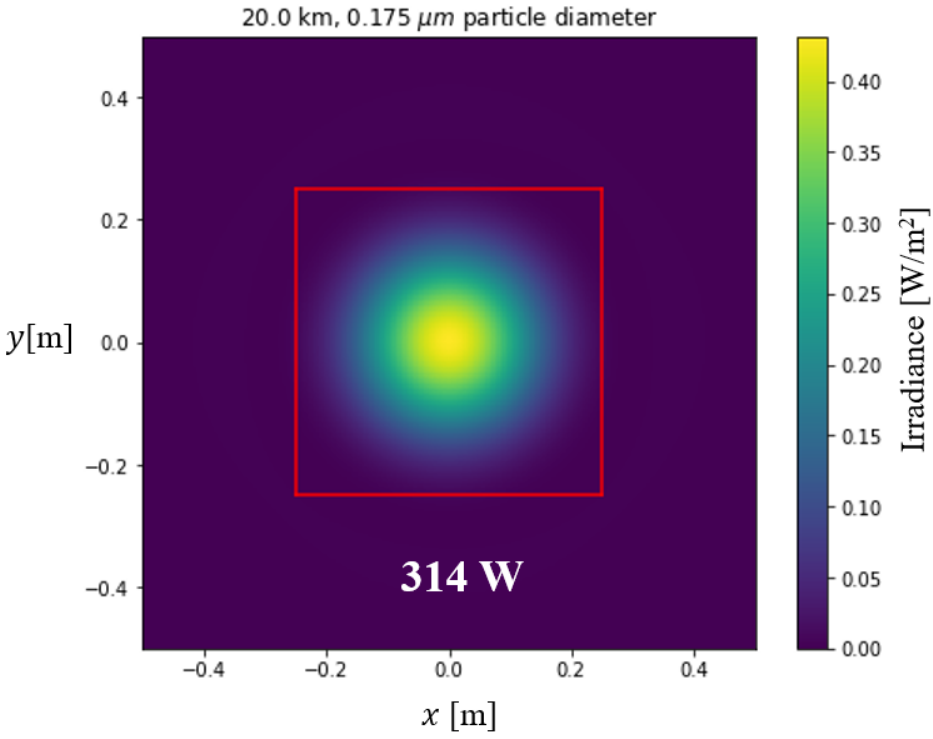}
	 }
\hfill
\subfloat[]{
	\label{subfig:20km_025um}
	\includegraphics[clip, scale=0.5]{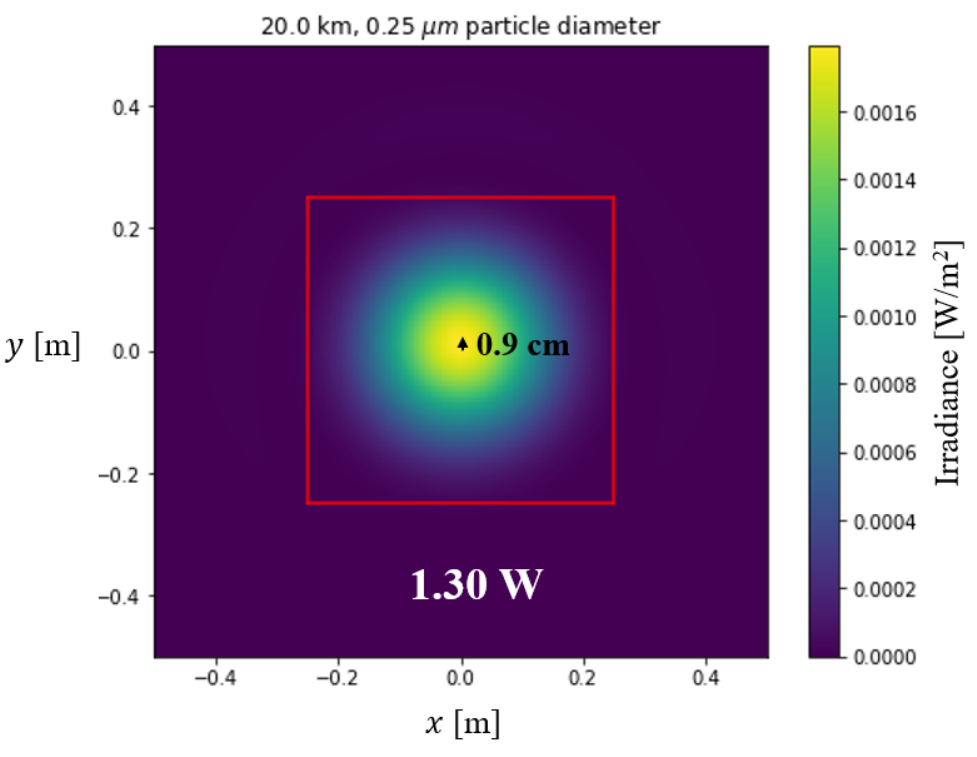}
	 }
     
\medskip
\subfloat[]{
	\label{subfig:50km_0175um}
	\includegraphics[clip, scale=0.5]{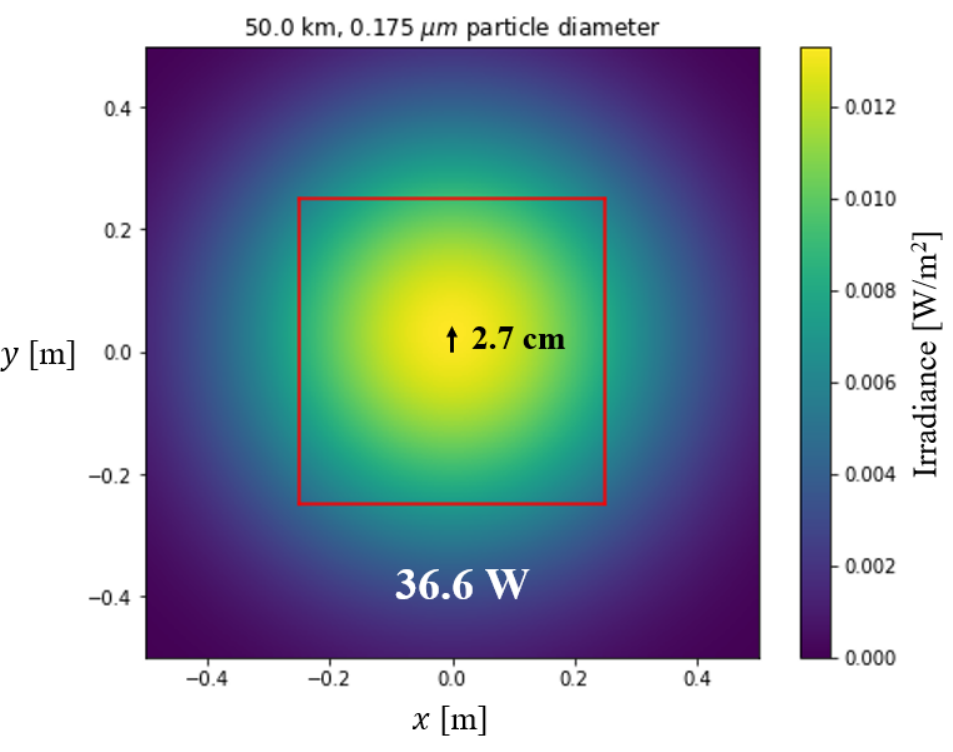}
	 }
\hfill
\subfloat[]{
	\label{subfig:50km_025um}
	\includegraphics[clip, scale=0.5]{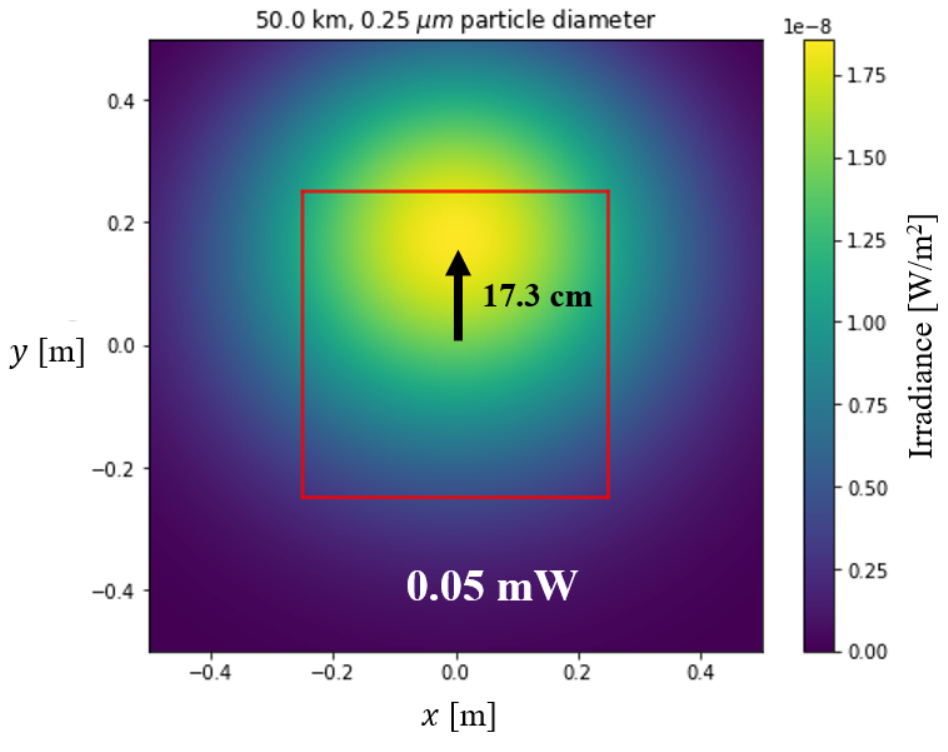}
	 }
\caption{Irradiance of panel for different distances and particle diameter sizes, with a laser source at 2 meters high. The arrow indicates the beam center shift on the panel due to non-uniform absorption.}
\label{fig:IrradiancePatterns}
\vspace{-0.15cm}
\end{figure*}

In the absence of lunar particles (“Beam diffraction (No Dust)” in Fig. \ref{fig:CenterToCenter_NoDust_Dust}), the loss of transferred power is caused by the gaussian diffraction as described by equation \ref{eq:irradiance_freespace}, where the beam radius increases causing it to irradiate outside the limits of the solar array. In such case, the efficiency starts to degrade at 25 km, going from $92.4\%$ to $50.4\%$ at 50 km. When considering dust (“Beam diffraction (With Dust)” in Fig. \ref{fig:CenterToCenter_NoDust_Dust}), the efficiency is reduced, and the significance of that reduction increases with length. Due to the high non-linearity of the particle concentration distribution in height, the loss cannot be simply approximated by considering propagation from the center of the laser to the center of the panel (“Center-to-center in dust” in Fig. \ref{fig:CenterToCenter_NoDust_Dust}) and assuming constant losses over the whole beam. Indeed, when comparing this simplistic assumption with the generalized diffraction model (“Beam diffraction (With Dust)” in Fig. \ref{fig:CenterToCenter_NoDust_Dust}), we notice a considerable decrease in efficiency at any distances for particles of 175 nm and for distances below 20 km for larger particles of 250 nm. This difference is not caused by a larger beam from Eq. \ref{eq:irradiance_freespace}, since the beam diffraction without dust (“Beam diffraction (No Dust)” in Fig. \ref{fig:CenterToCenter_NoDust_Dust}), has little effect on efficiency below 20 km, as can be observed by the beams of Fig. \ref{subfig:5km_0175um} and \ref{subfig:20km_0175um} remaining well within the panel. Rather, the difference is caused by a portion of the beam (the lower part) experiences higher losses than the center part of the beam added by the refraction effect of deviating the beam towards the high-loss region. Therefore, to get an adequate picture of the actual losses, considering a generalized diffraction model is critical.

The detrimental effect of dust can be mitigated by positioning the fixed emitter at a higher altitude. To optimize this choice, we show the efficiency with transmitter height and rover distance in Fig. \ref{fig:PowerTransfer_2D_plot} for two different particle diameters. Contrary to previous analysis \cite{Naqbi2024Impact}, the receiver’s height, i.e. the height of the solar panel on a rover, is fixed at 2 m, thus always inside the dust cloud. The choice of emitter altitude is therefore a compromise between efficiency and cost. Note that the scales are different for the two graphs to make them clearer to read. The power efficiency increases with the height of the laser simply because the dust cloud is less dense as we get higher and therefore beam is less exposed to particle scattering. For short distances, the impact of the height of the laser typically becomes less important as the emitter goes above the dust cloud. 

While, the beam will always have to cross the dust cloud at some point to reach the receiver, the distance traveled in the cloud is reduced as the horizontal angle between emitter and receiver is increased. In the case of particles with a diameter of 0.175 $\mu$m, for a laser at a height of 6, 9 and 12 meters, the efficiency of transfer for a distance between source and panel of 5 km are respectively $84\%$, $89\%$ and $91\%$ for 175 nm particle size. The effect of “being above the cloud” becomes less significant when powerbeaming to longer distances, as the path traveling inside the cloud remains significant and will incur losses. Not surprisingly, the efficiency drops with distance, due to beam diffraction widening beyond the solar panel and scattering losses. For instance, for a source placed at a height of 12 meters, the efficiencies of power transfer for distances of 5, 20 and 50 km are $91 \%$, $71\%$ and $25\%$ respectively for 175 nm particle size. These results highlight how the height placement of the transmitting source can improve transferred power. This applies as well to the placement of the solar panel, as we show in Fig. \ref{fig:PowervsHeightPanel}. Deploying a rover with a panel placed above the dust cloud (8.68 m high) is not realistic, but the results highlight how placing the array as high as possible can facilitate energy transfer. 

It is worth noticing that increasing the suspended particle diameter by about $42\%$ (from 175 nm to 250 nm), the situation gets significantly worse, as shown in Fig. \ref{subfig:2d025}. In such case, the system’s operating range is limited to 33 km, at which point we can achieve only $5 \%$ efficiency with a laser at 12 m in height, beyond which the efficiency is negligible. For a laser at a height of 6 m, the $5\%$ operating range is 16 km. For both heights, we could get beyond 50 km range at this efficiency for 175 nm particle. This is caused by the exponential increase in scattering cross-section with particle size. This $42\%$ size increase is equivalent to a 7 times increase in particle density. When the laser is placed 2 meters above the ground, the power transfer is around $19\%$ at 5 km and reaches around $5 \cdot 10^{-6}~ \%$ at 50 km. This particle size increase or 7 times concentration increase should not be considered improbable, since human activity in a lunar region may significantly increase suspended particle density and the size of the suspended particles. Fig. \ref{fig:PowervsParticle} displays how bigger particles (or higher densities) would make it impossible for sufficient power transfer, even at short distances. This highlights how the knowledge of particle size is critical to evaluate power transfer capacity. Furthermore, this indicates how human activities, causing the density of lunar dust or the number of bigger suspended particles to rise, impact power beaming. 

To fully understand and visualize the effect on non-uniform scattering on the beam, we display in Fig. \ref{fig:IrradiancePatterns} the irradiance pattern on the solar panel for fixed height of the laser and distance between source and the panel. The panel itself is represented by the red square. Only the irradiated region within the limits of the red square is counted when computing the received power. The value of the received power is displayed in red under each red square. One important thing to note is that the scales of irradiance for all subplots are different. This choice was made to be able to clearly see the gaussian pattern of the beam.

For the patterns with the same particle diameter, we observe that the received power decreases as the distance between the source and the panel increases. This is due in part to the scattering of the beam by the dust, but also, especially for long-distance such as 50 km, to beam widening outside the panel. As we look at Fig. \ref{subfig:20km_025um}, \ref{subfig:50km_0175um} and \ref{subfig:50km_025um}, we notice that the center of the Gaussian has shifted upward. At 20 km and particle diameter of 0.25 $\mu$m the shift is calculated to be 0.9 cm but cannot be clearly observed when looking only at the pattern \ref{subfig:20km_025um}. However, at 50 km, for 0.175 $\mu$m and 0.25 $\mu$m the calculated shift is respectively 2.7 cm and 17.3 cm and can be clearly seen on Fig. \ref{subfig:50km_0175um} and \ref{subfig:50km_025um}. This shifting of the beam over long distances is caused by the higher loss in the lower part of the beam due to higher particle concentration. It highlights that a small difference in the density of dust particles can cause significant differences for paths of light traveling at heights centimeters apart.

The results presented in this section reveal how scattering and absorption caused by lunar dust lead to significant power loss and beam distortion, especially over long transmission distances. The study further demonstrates that the efficiency of power delivery is highly sensitive to the size of dust particles, as well as the height of the laser source above the ground level. This emphasizes the importance of considering realistic dust models when assessing optical power transfer system on the surface of the Moon and designing devices with elevation strategies to achieve robust and efficient wireless power transfer for future lunar missions.

%% file: Sections/Conclusion.tex
This study had the objective to show the effect of lunar dust in a ground-to-ground WPT system in which a high-power laser placed on a tower at the surface of the Moon sends energy to a solar panel of a rover kilometers away. These results can be used to select the required height of the emitter to ensure a target efficiency. This was achieved using mathematical diffraction and lunar particle models that describe the propagation of the laser through the dust cloud of varying density. Simulations were conducted under various parameters such as the distance between the source and the rover, the height of the laser on the tower and the size of the particle’s diameter. Results highlight how placing the laser higher can increase the transferred power. Results also show how bigger particles can considerably reduce the amount of energy that can reach the panel, especially when the tower and the rover are far apart. 

This work can serve as a tool to calculate the feasibility of certain mission by giving the opportunity to adjust many parameters to compute the possible harvested power in complex situations.
While the larger suspended particles studied here exceed the expected average size of suspended particles under the sun’s exposure with our current limited knowledge of the moon, it is important to consider pessimistic scenarios which may become very probable following human activities on the moon. Such activity may significantly increase the dust density, particle size and suspension height. Our study shows that any slight increase in suspended particle size may have significant, if not mission-critical, impact on power-transfer efficiency. 

The accurate modeling  of Moon dust requires more data  but factors like the density of Moon dust and the size of particles in suspension above the surface can vary very quickly as soon as there is activity on the ground. A flexible model such as the one presented here can account for varying data will therefore remain relevant.
 

%% file: Biography.tex
\begin{IEEEbiography}
[{\includegraphics[width=1in,height=1.25in,clip,keepaspectratio]{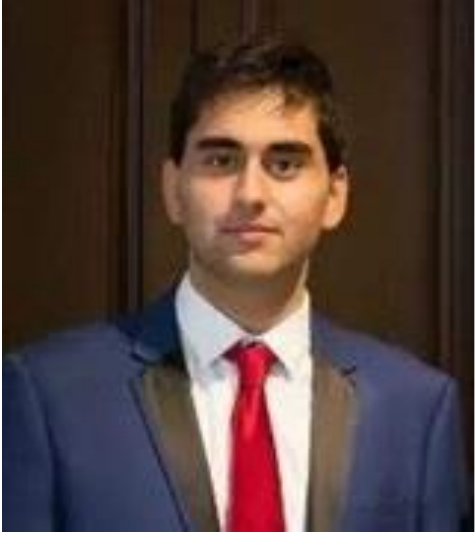}}] 
{Yanni Jiwan-Mercier}{\space} received his B. Eng.
 in physics engineering in 2020 and his M. Eng.
 in aerospace engineering in 2023, from École
 Polytechnique Montréal, Québec, Canada. He
 is currently doing his PH.D. degree in electrical
 engineering at Polytechnique Montréal on the
 subject of power transfer on the Moon.
\end{IEEEbiography}

\begin{IEEEbiography}
[{\includegraphics[width=1in,height=1.25in,clip,keepaspectratio]{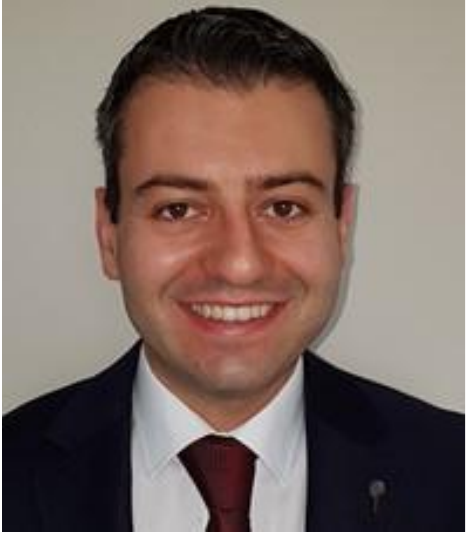}}] {Barış Dönmez}{\space}(Graduate Student Member, IEEE) received the B.Sc. and M.Sc. degrees (with high honors) in electrical and electronics engineering, in 2009 and 2022, respectively, from FMV Işık University, Istanbul, Turkiye. He is currently working toward a Ph.D. degree in electrical engineering at Polytechnique Montréal, Montréal, QC, Canada. His research interests include communication and energy harvesting systems in space networks.
\end{IEEEbiography}

\begin{IEEEbiography}
[{\includegraphics[width=1in,height=1.25in,clip,keepaspectratio]{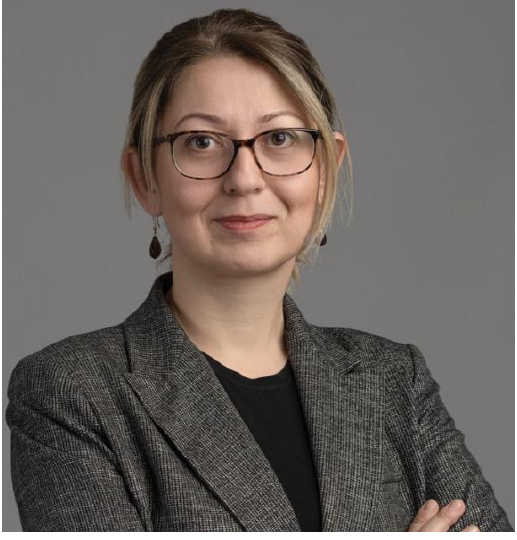}}] {Güneş Karabulut Kurt}{\space}(Senior Member, IEEE) is a Canada Research Chair (Tier 1) in New Frontiers in Space Communications and Associate Professor at Polytechnique Montréal, Montréal, QC, Canada. She is also an adjunct research professor at Carleton University, ON, Canada. Gunes is a Marie Curie Fellow and has received the Turkish Academy of Sciences Outstanding Young Scientist (TÜBA-GEBIP) Award in 2019. She received her Ph.D. degree in electrical engineering from the University of Ottawa, ON, Canada. 
\end{IEEEbiography}

\begin{IEEEbiography}
[{\includegraphics[width=1in,height=1.25in,clip,keepaspectratio]{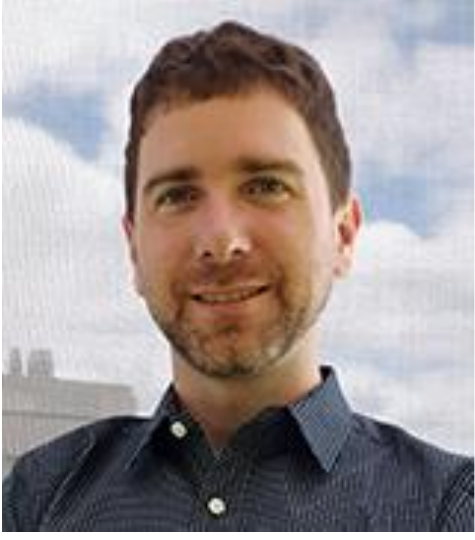}}] 
{Sébastien~Loranger}{\space}
 (Member, IEEE) is a
 Assistant Professor at Polytechnique Montreal,
 Montréal, Qc, Canada. He received Ph.D.
 degree in engineering physics in Polytechnique Montreal in 2018 and completed a post
doctoral fellowship at Max Planck Institute for
 the Science of Light, Erlangen, Germany in
 2020. His research focus is in photonics and
 optoelectronic components for space.
\end{IEEEbiography}

%% file: MAIN.bbl
% Generated by IEEEtran.bst, version: 1.14 (2015/08/26)
\begin{thebibliography}{10}
\providecommand{\url}[1]{#1}
\csname url@samestyle\endcsname
\providecommand{\newblock}{\relax}
\providecommand{\bibinfo}[2]{#2}
\providecommand{\BIBentrySTDinterwordspacing}{\spaceskip=0pt\relax}
\providecommand{\BIBentryALTinterwordstretchfactor}{4}
\providecommand{\BIBentryALTinterwordspacing}{\spaceskip=\fontdimen2\font plus
\BIBentryALTinterwordstretchfactor\fontdimen3\font minus \fontdimen4\font\relax}
\providecommand{\BIBforeignlanguage}[2]{{%
\expandafter\ifx\csname l@#1\endcsname\relax
\typeout{** WARNING: IEEEtran.bst: No hyphenation pattern has been}%
\typeout{** loaded for the language `#1'. Using the pattern for}%
\typeout{** the default language instead.}%
\else
\language=\csname l@#1\endcsname
\fi
#2}}
\providecommand{\BIBdecl}{\relax}
\BIBdecl

\bibitem{nasa_artemis}
\BIBentryALTinterwordspacing
{NASA}, ``Artemis program overview,'' 2020, {Access Date:} 7.7.2025. [Online]. Available: \url{https://www.nasa.gov/specials/artemis/}
\BIBentrySTDinterwordspacing

\bibitem{esa_moon}
\BIBentryALTinterwordspacing
{ESA}, ``{ESA}'s moon village concept,'' 2021, {Access Date:} 7.7.2025. [Online]. Available: \url{https://www.esa.int/About_Us/Moon_Village}
\BIBentrySTDinterwordspacing

\bibitem{private_moon}
{SpaceX, Blue Origin, and others}, ``Private sector involvement in lunar exploration,'' 2022, general industry news and whitepapers.

\bibitem{nasapowerneeds}
W.~M. Farrell, T.~J. Stubbs, and R.~R. Vondrak, ``Exploring lunar electrical environment and its influence on surface systems,'' \emph{Space Weather}, vol.~6, no.~4, p. S04, 2008.

\bibitem{donmez2024continuous}
\BIBentryALTinterwordspacing
B.~Donmez and G.~{Karabulut Kurt}, ``Continuous power beaming to lunar far side from {EMLP}-2 halo orbit,'' 2024, {Access Date:} 7.7.2025. [Online]. Available: \url{https://arxiv.org/abs/2402.16320}
\BIBentrySTDinterwordspacing

\bibitem{naqbi2024opticalpowerbeaminglunar}
M.~Naqbi, S.~Loranger, and G.~K. Kurt, ``Optical power beaming in the lunar environment,'' \emph{IEEE Transactions on Aerospace and Electronic Systems}, 2025.

\bibitem{Stubbs2006Dynamic}
T.~J. Stubbs, R.R.V., and W.~M. Farrell, ``{A dynamic fountain model for lunar dust},'' \emph{{Advances in Space Research}}, vol.~37, 2006.

\bibitem{horanyi2015}
M.~e.~a. Horányi, ``A permanent, asymmetric dust cloud around the moon,'' \emph{Nature}, vol. 522, p. 324–326, 2015.

\bibitem{szalay2016}
J.~R. Szalay and M.~Horányi, ``Lunar meteoritic ejecta cloud: Density distribution and mass flux,'' \emph{Geophysical Research Letters}, vol.~43, no.~10, pp. 4893--4898, 2016.

\bibitem{bozek1995laserbeam}
\BIBentryALTinterwordspacing
J.~M. Bozek, ``Ground-based and space-based laser beam power applications,'' National Aeronautics and Space Administration, Lewis Research Center, Cleveland, OH, NASA Technical Memorandum 106744, Feb. 1995, accessed: 2025-07-09. [Online]. Available: \url{https://ntrs.nasa.gov/api/citations/19950016253/downloads/19950016253.pdf}
\BIBentrySTDinterwordspacing

\bibitem{donmez2025multiorbitercontinuouslunarbeaming}
\BIBentryALTinterwordspacing
B.~Donmez, Y.~Jiwan-Mercier, S.~Loranger, and G.~{Karabulut Kurt}, ``Multi-orbiter continuous lunar beaming,'' 2025. [Online]. Available: \url{https://arxiv.org/abs/2504.11300}
\BIBentrySTDinterwordspacing

\bibitem{donmez2025hybrid}
B.~Dönmez, Y.~Jiwan-Mercier, S.~Loranger, and G.~K. Kurt, ``Hybrid {FSO} and {RF} lunar wireless power transfer,'' in \emph{Proceedings of the 18th International Conference on Space Operations (SpaceOps 2025)}.\hskip 1em plus 0.5em minus 0.4em\relax Montreal, Canada: Canadian Space Agency on behalf of SpaceOps, May 2025, pp. 1--12, paper ID: 446.

\bibitem{VoltaSpace}
\BIBentryALTinterwordspacing
{Volta Space Technologies}, ``\BIBforeignlanguage{en-US}{Powering our first lunar rover},'' {Access Date:} 07.07.2025. [Online]. Available: \url{https://www.voltaspace.co/demonstrations}
\BIBentrySTDinterwordspacing

\bibitem{VanRiesen2001Gaas}
S.~Van~Riesen, U.S., and A.~W. Bett, ``{GaAs photovoltaic cells for laser power beaming at high power densities},'' in \emph{{Proc. Eur. PV Solar Energy Conference}}, 2001, pp. 18--21.

\bibitem{Raible2008High}
D.~Raible, ``{High Intensity Laser Power Beaming for Wireless Power Transmission},'' Ph.D. dissertation, Cleveland State University, Cleveland, OH, 2008.

\bibitem{He2014HighPower}
T.~He, S.Y., Z.~Changming, and M.~A. Munoz-Garcia, ``{High-Power High-Efficiency Laser Power Transmission at 100 m Using Optimized Multi-Cell GaAs Converter},'' \emph{{Chinese Optics Letters}}, 2014.

\bibitem{York2017High}
M.~York and S.~Fafard, ``{High efficiency phototransducers based on a novel vertical epitaxial heterostructure architecture (VEHSA) with thin p/n junctions},'' \emph{{Journal of Physics D: Applied Physics}}, vol.~50, no.~17, p. 173003, 2017.

\bibitem{Cavallaro2019Researchers}
E.~Cavallaro, ``{Researchers transmit energy with laser in ‘historic’ power-beaming demonstration},'' 2019, {U.S. Naval Laboratory}.

\bibitem{Huang2015Recent}
R.~K. Huang, B.~Samson, B.~Chann, B.~Lochman, and P.~Tayebati, ``Recent progress on high-brightness k{W}-class direct diode lasers,'' in \emph{2015 IEEE High Power Diode Lasers and Systems Conference (HPD)}.\hskip 1em plus 0.5em minus 0.4em\relax IEEE, 2015, pp. 29--30.

\bibitem{Platonov2020HighEfficient}
N.~Platonov \emph{et~al.}, ``{High-efficient kW-level single-mode ytterbium fiber lasers in all-fiber format with diffraction-limited beam at wavelengths in 1000-1030 nm spectral range},'' in \emph{{SPIE LASE}}, vol. 11260, 2020.

\bibitem{IPG2025High}
\BIBentryALTinterwordspacing
{IPG Photonics}, ``{High power fiber lasers},'' 2025, {Access Date:} 7.7.2025. [Online]. Available: \url{https://www.ipgphotonics.com/products/lasers/industrial-cw-fiber-lasers/high-power-fiber-lasers}
\BIBentrySTDinterwordspacing

\bibitem{Nikolay2020Optimization}
A.~Nikolay \emph{et~al.}, ``{Optimization of photoelectric parameters of InGaAs metamorphic laser ($\lambda$=1064 nm) power converters with over 50\% efficiency},'' \emph{{Solar Energy Materials and Solar Cells}}, vol. 217, p. 110710, 2020.

\bibitem{Helmers202168}
H.~Helmers \emph{et~al.}, ``{68.9\% Efficient GaAs-Based Photonic Power Conversion Enabled by Photon Recycling and Optical Resonance},'' \emph{{Rapid Research Letters}(RRL)}, vol.~15, no.~7, p. 2100113, 2021.

\bibitem{Fafard202274}
S.~Fafard and D.~Masson, ``{74.7\% Efficient GaAs-Based Laser Power Converters at 808 nm at 150 K},'' \emph{{Photonics}}, vol.~9, no.~8, p. 579, 2022.

\bibitem{USNRL2023First}
{U.S.N.R. Laboratory}, ``{First In-Space Laser Power Beaming Experiment Surpasses 100 Days of Successful Operation},'' 2023.

\bibitem{Naqbi2024Impact}
M.~Naqbi, S.~Loranger, and G.~Karabulut~Kurt, ``{Impact of Lunar Dust on Free Space Optical (FSO) Energy Harvesting},'' in \emph{{IEEE Aerospace Conference}}, Big Sky, MT, USA, 2024, pp. 1--9.

\bibitem{Rennilson1974Surveyor}
R.~Rennilson and D.~Criswell, ``{Surveyor observations of lunar horizon-glow},'' \emph{{The Moon}}, vol.~10, no.~2, pp. 121--142, 1974.

\bibitem{Colwell2007Lunar}
J.~E. Colwell, S.B., M.~Horanyi, S.~Robertson, and S.~Sture, ``{Lunar Surface: Dust Dynamics and Regolith Mechanics},'' \emph{{Review of Geophysics}}, vol.~45, no.~2, 2007.

\bibitem{Poppe2010Simulations}
A.~Poppe and M.~Horányi, ``{Simulations of the photoelectron sheath and dust levitation on the lunar surface},'' \emph{{Journal of Geophysical Research: Space Physics}}, vol. 115, p. A08106, 2010.

\bibitem{Szalay2015Search}
J.~Szalay and M.~Horányi, ``{The search for electrostatically lofted grains above the Moon with the Lunar Dust Experiment},'' \emph{{Geophysical Research Letters}}, vol.~42, no.~13, pp. 5141--5146, 2015.

\bibitem{Agui2007Lunar}
J.~Agui, ``{Lunar Dust Characterization for Exploration Life Support Systems},'' NASA Glenn Research Center, Tech. Rep., 2007.

\bibitem{Vidwans2022Size}
A.~Vidwans \emph{et~al.}, ``{Size and charge distribution characteristics of fine and ultrafine particles in simulated lunar dust: Relevance to lunar missions and exploration},'' \emph{{Planetary and Space Science}}, vol. 210, p. 105392, 2022.

\bibitem{Hartzell2011Role}
C.~M. Hartzell and D.~J. Scheeres, ``The role of cohesive forces in particle launching on the moon and asteroids,'' \emph{Planetary and Space Science}, vol.~59, no.~14, pp. 1758--1768, 2011.

\bibitem{Gaier2005Effects}
J.~Gaier, ``{The Effects of Lunar Dust on EVA Systems during the Apollo Missions},'' Glenn Research Center, Cleveland, Ohio, Tech. Rep., 2005.

\bibitem{Taylor2001Lunar}
L.~Taylor \emph{et~al.}, ``{Lunar Mare Soils: Space weathering and the major effects of surface-correlated nanophase Fe},'' \emph{{Journal of Geophysical Research: Planets}}, vol. 106, no. E11, pp. 27\,985--27\,999, 2001.

\bibitem{Szalay2016Lunar}
J.~Szalay and M.~Horányi, ``{Lunar meteoritic gardening rate derived from in situ LADEE/LDEX measurements},'' \emph{{Geophysical Research Letters}}, vol.~43, no.~10, pp. 4893--4898, 2016.

\bibitem{Horanyi2014Lunar}
M.~Horányi \emph{et~al.}, ``{The Lunar Dust Experiment (LDEX) Onboard the Lunar Atmosphere and Dust Environment Explorer (LADEE) Mission},'' \emph{{Space Science Reviews}}, vol. 185, no.~1, pp. 93--113, 2014.

\bibitem{Jin2024Properties}
H.~Jin \emph{et~al.}, ``{Properties of Lunar Dust and Their Migration on the Moon},'' \emph{{Space: Science \& Technology}}, vol.~4, p. 0142, 2024.

\bibitem{saleh2019fundamentals}
B.~E.~A. Saleh and M.~C. Teich, \emph{Fundamentals of Photonics}, 3rd~ed.\hskip 1em plus 0.5em minus 0.4em\relax Hoboken, NJ: John Wiley \& Sons, 2019.

\bibitem{donmezatp}
B.~Donmez, I.~Azam, and G.~{Karabulut Kurt}, ``Mitigation of misalignment errors over {Inter-Satellite} {FSO} energy harvesting,'' in \emph{2023 IEEE 34th Annual International Symposium on Personal, Indoor and Mobile Radio Communications (PIMRC)}.\hskip 1em plus 0.5em minus 0.4em\relax IEEE, 2023, pp. 1--5.

\bibitem{lasertypes}
H.~Kaushal and G.~Kaddoum, ``Optical communication in space: Challenges and mitigation techniques,'' \emph{IEEE Communications Surveys \& Tutorials}, vol.~19, no.~1, pp. 57--96, 2017.

\end{thebibliography}
